\documentclass[12pt,preprintnumbers]{article}
\usepackage{amsmath, amssymb, wasysym, slashed, multirow,hyperref}
\usepackage{pdfpages}
\usepackage{cite}
\usepackage{dsfont}
\usepackage{times}
\usepackage{xcolor}
\usepackage{graphicx,color}  %
\usepackage{bm}  %
\usepackage{enumitem}
\newenvironment{sciabstract}{%
\begin{quote} }
{\end{quote}}

\newcounter{lastnote}

\usepackage{fancyhdr}
\fancypagestyle{plain}{%
	\fancyhead[R]{}

}

\title{Quantum Field Theory on Multifractal Spacetime: Varying Dimension and Ultraviolet Completeness}

\author
{Alessio Maiezza$^{1,2\ast}$, Juan Carlos Vasquez$^{3,4\dagger}$\\
\\
\normalsize{$^{1}$Dipartimento di Scienze Fisiche e Chimiche, Universit\`a} \\ \\
\normalsize{degli Studi dell'Aquila, via Vetoio, I-67100, L'Aquila, Italy,}\\ \\
\normalsize{$^{2}$INFN, Laboratori Nazionali del Gran Sasso, 67010 Assergi, L'Aquila, Italy,}\\ \\
\normalsize{$^{3}$SISSA, Via Bonomea 265, I-34136 Trieste, Italy,} \\ \\
\normalsize{$^{4}$Centro de Fisica Fundamental, Universidad de Los Andes, Merida, Venezuela.} \\ \\
\small{ E-mail: alessiomaiezza@gmail.com$^{\ast}$, juancarlos8866@gmail.com$^{\dagger}$}
}

\date{}

\begin{document}


\baselineskip16pt 


\maketitle


\begin{sciabstract}
Inspired by various quantum gravity approaches, we explore quantum field theory where spacetime exhibits scaling properties and dimensional reduction with changing energy scales, effectively behaving as a multifractal manifold. Working within canonical quantization, we demonstrate how to properly quantize fields in such a multifractal spacetime. Our analysis reveals that a non-differentiable nature of spacetime is not merely compatible with quantum field theory but significantly enhances its mathematical foundation. Most notably, this approach ensures perturbative UV finiteness and improved behavior of the series expansion and enables rigorous construction of the $S$-matrix in the interaction picture by breaking vacuum translational invariance. The multifractal structure tames dominant, large-order divergence sources in the perturbative series and resolves the Landau pole problem through asymptotic safety, substantially improving the theory's behavior in the deep ultraviolet regime. Our formulation preserves all established predictions of standard quantum field theory at low energies while offering novel physical behaviors at high energy scales. 
\end{sciabstract}

\section{Introduction}\label{sec:introduction}

While the concept of dimension spans broadly across geometry, physics often simplifies it by stating that we inhabit a three-dimensional space. Our macroscopic experience confirms this, yet exceptions may emerge at the microscopic level. Mathematics offers various definitions of dimension -- the topological dimension $d$ represents an integer, with $d=3$ identified as the dimension of our physical space. In contrast, certain definitions do not require integer values, such as the Hausdorff dimension, which becomes essential for describing fractal manifolds \cite{mandelbrot1983fractal}.

Consider a sheet of paper, modeled topologically as two-dimensional. When crumpled, however, it assumes a different effective fractal dimension. We can estimate this physically by measuring the crumpled paper's density. Solid objects have densities scaling as $r^{-3}$ (where $r$ represents a characteristic length), yielding $d=3$ as a dimensionality estimate that coincides with the topological dimension. The crumpled paper, however, follows a different scaling relation, typically $r^{-d^{eff}}$, where $2<d^{eff}<3$.
This simple example illustrates a crucial insight: the physical dimension may be neither trivial nor externally fixed. Rather, it requires a ``physical estimator", a concept central to this work, analogous to the density in our example.

Such estimators become vital in quantum realms, where direct dimensional experience is impossible. Quantum field theory (QFT) provides a natural dimension estimator through the Green function. In Euclidean space, this function diverges as $r^{-(d-2)}$ at short distances. Conventionally, dimension is treated as a fixed external parameter conditioned by macroscopic experience. However, we can invert this logic: we examine Green function behavior at short distances to estimate the spacetime dimension \footnote{The concept that the dimension must be \textit{a posteriori} determined was introduced in \cite{Hochberg:1989tr}, specifically via a variational principle.}. This transforms dimension into an observable quantity that may differ from the topological dimension $d$, an approach we adopt throughout this paper.

How should dimension differ from large-scale observations? Space-time could possess intrinsic fractal properties, causing fields to experience a different dimension than the topological one. The idea of a rough spacetime was first proposed by Wheeler, who called it spacetime foam \cite{wheeler1963} -- see also \cite{Ng:2008pi,Carlip:2022pyh}. We shall broadly refer to the roughness (non-differentiability) of spacetime as ``fractal spacetime". While string theory often suggests topological dimensions exceeding four \footnote{Dimension reduction can also be formulated within string theory \cite{Afshordi:2014cia} and multibrane model \cite{Dai:2014roa}.}, many quantum gravity approaches propose the opposite: dimensional reduction. Theories including Causal Dynamical Triangulations \cite{Ambj_rn_2005,Kommu_2012,Coumbe_2015}, Asymptotic Safety \cite{einstein1979general,Lauscher_2001}, Causal Set Theory \cite{Carlip_2015}, and Loop Quantum Gravity \cite{Modesto_2009} suggest that effective dimension flows continuously from four at large scales to two at short distances. Additional perspectives appear in \cite{Benedetti_2009,Mondal_2022}. A model universe with variable dimension is presented in \cite{Mansouri_1999}. For clarity, let us stress that the term "(multi)fractal spacetime" in this work refers to an effective property derived from QFT propagator scaling, discussed below, and not to multifractional geometries \cite{Calcagni:2010fractal,Calcagni:2012multifractional,Calcagni:2012geometry,Calcagni:2013multifractal,Calcagni:2017review,Calcagni:2021aap,Calcagni:2022shb}. We emphasize that we work with a flat, fractal space time, motivated from quantum gravity arguments, as discussed above.

This article examines this phenomenon from the QFT perspective, broadening the seminal work of \cite{Svozil:1985ha} on QFT formulation in fractal spacetime in many directions.  The propagator, central to QFT, can serve as a dimensional estimator through running couplings. We investigate how canonical quantization should be modified to accommodate fractal space-time and dimensional reduction. Our findings reveal a straightforward generalization of canonical quantization that significantly enhances QFT consistency. First, the theory becomes finite at all orders in perturbation theory. Second, the behavior of the perturbative series is improved since the leading singularities in the Borel plane, namely, the ultraviolet (UV) renormalons, disappear. Third, the UV Landau pole also disappears so that any QFT model embedded in multifractal spacetime becomes asymptotically safe. Finally, a remarkable feature is the Poincaré non-invariance of the vacuum, enabling robust $S$-matrix construction by circumventing limitations imposed by the Haag theorem \cite{Haag:1955ev}. 

While form factors, damping UV behavior, have been extensively studied in nonlocal quantum field theory and quantum gravity, motivated by string field theory and the need to control ultraviolet divergences while preserving unitarity and ghost-freeness \cite{Krasnikov:1987,Kuzmin:1989,Tomboulis:1997,Modesto:2012prd,Biswas:2012prl,Modesto:2015,Modesto:2017review,BasiBeneito:2022review,Buoninfante:2022review,Calcagni:2023spectral}, here we modify the propagator, via a function (form factor) that would break the translational invariance of the vacuum, on which Haag's theorem relies. Hence, enabling a consistent perturbative framework beyond what standard nonlocal models achieve. 

In Sec. \ref{Reduction}, we provide a self-contained analysis of how multifractality may be connected to dimensional reduction, consistent with the quantum gravity approaches mentioned above. We address canonical quantization in Sec. \ref{sec:Quantization}, then we discuss the consistency in Sec. \ref{Consistency}, and the implications in Sec. \ref{Discussion}. We give our outlook in Sec. \ref{Summary}. Additional technical details are reported in the appendix \ref{HT}.

\section{Fractality, smoothing and dimensional reduction   }\label{Reduction}


This preliminary section aims to connect the concepts of multifractal space and varying effective dimensions. 

To start with, consider first a 1D space where the tangent is not defined at any point, namely, is not smooth, and consider a non-differentiable function $f(x):\mathcal{M} \rightarrow \mathbb{R}$ on this 1D manifold $\mathcal{M}$. It is possible to build from $f$ a new function $F:\mathcal{M} \rightarrow \mathbb{R}$, differentiable, via convolution with an auxiliary smoothing function $h(x)$ that can be chosen Gaussian\cite{Nottale:1989ag}, 
\begin{equation}
h(x)= \frac{2}{\sqrt{4 \pi l^2}} e^{-\frac{x^2}{4 l^2}}\,.
\end{equation}
The convolution product $f*h$ gives
\begin{equation}\label{convolution}
F(x,l)=  \int_{0}^{\infty} f(y) \, h(x-y)\, dy \,,
\end{equation}
now being $F$ smoothed, differentiable, and dependent on an intrinsic scale $l$, implying resolution dependence. By construction, when $l\rightarrow 0$, the function $F$ coincides with the original non-differentiable function $f(x)$. 

For instance, think of the above function $f(x)$ as the distance between a horizontal line and a point in the 1D non-differentiable line given by the points belonging to the Koch curve.  The function $f$ is continuous everywhere in $x$ but non-differentiable anywhere in $x$ \cite{von_koch_1904}. Now, assume that one is interested in finding the distance between any two points in the Koch curve. For zero resolution, such a distance is not well-defined (infinite) when $l\rightarrow 0$, but it does exist for a finite, nonzero resolution $l$. After applying \eqref{convolution}, the distance between two points in the Koch curve can be estimated and is given by 
\begin{equation}
 d(l) = \int_a^b \, \sqrt{1+(\partial_x F(x,l)})^2\, dx \,,
\end{equation}
where $a$ and $b$ are the $x$-coordinates of the two points in the Koch curve,  and $\partial_x$ denotes derivation with respect of $x$.
 
The length of a segment on the Koch curve depends on the smoothing function $h$ and the scale parameter $l$. We shall refer to this interplay between the resolution and non-differentiability as ``fractality" in a broad sense. In Sec.\ref{sec:Quantization},  we shall show that something similar can happen in QFT, where the role of $f$ and $F$ is played by the fields before and after smoothing, respectively,  and the role of $d(l)$ is played by the action functional.

\subsection{Introducing a scale in Euclidean Green equation: Dimensional Reduction}\label{sec:DR}

Let us now move to the case of a $\mathbb{R}^d$ non-smooth, Euclidean manifold. Similarly to the function $F$ in \eqref{convolution}, the Green function has to depend on a scale that we call $M$, with the dimension of a mass-energy. The physical interpretation is that the space appears smooth at energy much below $M$, but it starts to show its fractal features at higher energy, comparable with $M$. This may be either an intrinsic scale of Nature or an effective energy emerging from a deeper theory than QFT, a theory that should include gravity at the quantum level \footnote{Fractal properties induced by quantum gravity effects are discussed in \cite{Nicolini_2011}.}.
Our approach  provides an \emph{operational definition} of an effective dimension directly from the QFT propagator, distinct from the usual spectral dimension.

The standard Green for the Klein-Gordon equation,
\begin{equation}
\left(\partial_\mu \partial^\mu  - m^2 \right) G(x-y)= -\delta^{(d)}(x-y) \,,
\end{equation}
becomes
\begin{equation}\label{newG}
\left(\partial_\mu \partial^\mu  - m^2 \right) G(x-y, M)= -g(x-y,M) \,,
\end{equation}
where $g(k,M)$, represents the smoothing of the Dirac's delta since the left-handed side of \eqref{newG} now depends on $M$. The standard case is reproduced for $M\rightarrow\infty$. For simplicity, we choose $g$ in the Gaussian form,
\begin{equation}\label{smoothing_function}
g(x,M):= \left( \frac{M^2}{4\pi}  \right)^\frac{d}{2} \, e^{-\frac{1}{4} M^2 x^2}  \,.   
\end{equation}
The replacement of $\delta^{(d)}(x-y)$ with a Gaussian kernel  is consistent with the Schwinger proper-time representation and heat-kernel techniques widely used in QFT \cite{Vassilevich:2003,Barvinsky:1985,Buchbinder:2021book}. Form factors in nonlocal theories implement similar exponential suppression to achieve UV control without violating gauge invariance. In the framework discussed in this work, the Gaussian kernel models a resolution scale $M$, signaling the onset of multifractal features rather than imposing a hard cutoff in the theory.

The Fourier transform of \ref{newG} reads,
\begin{equation}\label{newGp}
\left(p^2+m^2\right) G(p,M) = e^{-\frac{p^2}{M^2}}   \,,
\end{equation}
leading to
\begin{equation}\label{new_prop}
G(p,M)= \frac{e^{-\frac{p^2}{M^2}}}{p^2+m^2}  \,.
\end{equation}
Next, let us do the inverse-Fourier transform of \eqref{new_prop} in $O_d$ symmetry and in the massless limit, 
\begin{equation}\label{int}
G(s,M)= 2^{d-2} \pi ^{d/2} s^{2-d} \left(\Gamma \left(\frac{d}{2}-1\right)-\Gamma \left(\frac{d}{2}-1,\frac{M^2
   s^2}{4}\right)\right)  \,,
\end{equation}
being $s=\sum_i^d x_i^2$, $\Gamma(x)$ the gamma function, and $\Gamma(a,x)$ the incomplete gamma function. 

For the topological dimension $d>2$ and $M\rightarrow\infty$, one knows that the standard result, $G(s)$, diverges as $s^{2-d}$ for $s\rightarrow 0$. In contrast, the result in \eqref{int} is finite as $s\rightarrow 0$. We want to encode this degree of divergence or convergence into an estimator of the effective dimension. To this aim, it is convenient to define
$R:= s^{-\frac{2-d}{d}}$, such that $G(R)\sim R^d$, and thus the logarithm derivative of $G$ gives,
\begin{equation}
\frac{d \log(G(R))}{d\log R}=d \,.
\end{equation}
This equation assigns \textit{ad hoc} an effective dimension $d$, equivalent to the topological dimension, in the standard case. 
The reason is to define the effective dimension by replacing $G(R) \rightarrow G(R,M)$, 
\begin{equation}\label{gendeff}
d^{eff}:= \frac{d \log(G(R,M))}{d\log R} = d-\frac{2^{3-d} d e^{-\frac{1}{4} M^2 s^2} (M s)^{d-2}}{(d-2) \left(\Gamma \left(\frac{d}{2}-1\right)-\Gamma
   \left(\frac{d}{2}-1,\frac{M^2 s^2}{4}\right)\right)}   \,,
\end{equation}
where in the last step we have rewritten the result in terms of the original variable, $s$.

The expression in \eqref{gendeff} greatly simplifies in the phenomenological case for which the topological dimension is four, $d=4$,
\begin{equation}\label{4deff}
d^{eff}|_{d=4} = 4-\frac{M^2 s^2}{e^{\frac{M^2 s^2}{4}}-1}  \,. 
\end{equation}
Therefore, $d^{eff}$ continuously flows from 4, at low energy, to $d^{eff}=2$ at $r \approx 2.2 M^{-1}$, and it asymptotically reaches $d^{eff}=0$ at infinite energy. This dimensional reduction from 4 to 2 is consistent with findings from various quantum gravity approaches. The spectral dimension, defined via heat-kernel diffusion on quantum spacetimes \cite{Ambjorn:2005,Carlip:2017eud,Eckstein:2020gjd}, provides a measure of effective dimensionality. Our operational definition, based on the propagator scaling, differs from but complements the spectral dimension approach. The interpretation is that $d^{eff}$ is an energy-dependent estimator of the non-constant dimension of the space. Specifically, the quantity $d^{eff}$ may be interpreted as an estimator of the  ``singularity spectrum" defining spacetime as a multi-fractal manifold \cite{harte2001multifractals,falconer2013fractal} -- see \cite{Calcagni_2010,Calcagni:2010bj} for other applications to physical models. It is worth stressing that for energy close to $M$, the effective dimension becomes two, in agreement with many, if not all, suggestions from approaches to quantum gravity -- see \cite{Carlip_2017} for a comprehensive review. At energy much larger than $M$, the effective dimension vanishes. At energies much larger than the Planck energy, the effective dimension vanishes, which may not be immediately apparent. However, if we assume that $M$ is comparable to the Planck energy, we can speculate that the space-time dimensionality loses its meaning at energies where quantum fluctuations of space-time become dominant. This aligns with the analysis based on the spectral dimension estimator \cite{Modesto_2010}.

Before attempting the canonical quantization in this scenario, in the next section, let us summarize and comment on the picture we have introduced. First, the spacetime has topological dimension $d$ (e.g., $d=4$), thus one calculates in the standard way any algebra, vector, tensor, etc., usually appearing in QFT. Second, we conceptualize that quantum fields perceive an effective (varying) dimension $d^{eff}<d$, and this is defined by the behavior of the propagator for large momenta. The highlighted scenario shares some features with the one in \cite{Svozil:1985ha}, being one main difference that we consider and motivate a varying effective dimension. Another distinction is that we face the problem of rendering consistent canonical quantization in QFT, thereby dealing with well-known constraints. 

{It is worth noting that nonlocal QFT and gravity models with damping  functions have been developed in depth, addressing unitarity, super-renormalizability, and in some cases finiteness \cite{Modesto:2012prd,Tomboulis:2015prd,Buoninfante:2020prd}. As we will see in the rest of this paper, our proposal differs conceptually by introducing a vacuum that is not translationally invariant, requiring a modification of the above form factor. In this way, we circumvent Haag's theorem and establish a perturbative framework that is both UV finite and free of renormalon ambiguities.

\section{Canonical quantization with varying effective dimension}\label{sec:Quantization}

We now turn our attention to actual QFT (in 1+3 topological dimensions). The aim is to generalize the canonical quantization of fields to the case of a fractal space-time with a running effective dimension.

Consider a real scalar field $\phi(x)$  with action 
\begin{equation}
    S = \int d^4x \,\left(\frac{1}{2}(\partial_{\mu}\phi(x))^2-\frac{1}{2}m^2 \phi(x)^2 -\frac{\lambda}{4!}\phi(x)^4\right) \,,
\end{equation}

and write the free Feynman propagator,
\begin{equation}\label{standard_Fey}
G_F(x-y) = \langle 0 | \mathcal{T} \phi(x) \phi(y)  | 0 \rangle  \,,
\end{equation}
with $\mathcal{T}$ being the time-ordering operator.

In the same spirit as section \ref{sec:DR}, assume that space-time has fractal properties at scale $M$ and thus $G_F(x-y)$ has to be smoothed as $G_F(x-y,M)$. From \eqref{standard_Fey}, it follows that the field $\phi$ must depend on the parameter $M$. One can visualize the field $\phi(x,M)$ in terms of a convolution with a smoothing function, parametrized by $M$, as in \eqref{convolution}. 

As in the standard QFT, it is convenient to write the classical field, before quantization, in the Fourier representation (fixing $d=4$)
\begin{equation}
\phi(x,M) = \frac{1}{(2\pi)^4} \int \phi(p,M) e^{i p x} d^4 p    \,,
\end{equation}
which satisfies the Klein-Gordon equation, which is in momentum space
\begin{equation}
 (p^2-m^2)\phi(p,M) =0   \,.
\end{equation}
The solution is of the form, 
\begin{equation} \label{eq:phi_p_classic}
 \phi(p,M) = 2\pi \, \delta(p^2-m^2) \, a(p,M) \,,
\end{equation}
where $\delta$ is the Dirac's delta function, and $a(p,M)$ is an arbitrary, non-singular function of the momentum and the parameter $M$, which we define as
\begin{equation}
a(p,M) := a(p) \,\sqrt{r(p,M)} \,,
\end{equation}
Then, the $M$-dependent field can be written as 
\begin{equation} \label{eq:phi_p_classic}
 \phi(p,M) = \phi(p)\, \sqrt{r(p,M)} \,.
\end{equation}
Notice that this approach is radically different than the approach adopted in \cite{Calcagni:2010bj,Maiezza:2022xqv}, where the classical equations of motion are modified by a non-propagating field. 

Essentially, the standard quantization consists in promoting $a(p,M)$ from a function to an operator, so defining the ladder operators with a dependence on the scale $M$.

The function $r$ is  positive defined, and it is such that $r \approx 1$ for $p \ll M$, reproducing standard QFT at low energy, but changing physics at deep UV. Then we have \footnote{To keep contact with \cite{Svozil:1985ha}, this expression can be regarded as a Stieltjes-Fourier transform of measure $d \mu_H:= \sqrt{r(p,M)} d^4p$.},
\begin{equation}
\phi(x,M) = \frac{1}{(2\pi)^4} \int \sqrt{r(p,M)} \phi(p) e^{i p x} d^4 p \,,
\end{equation}
which can be written explicitly as a non-local operator as:
\begin{equation}
\phi(x,M)  = \int d^4 y\, R(x-y,M)\, \phi(y)\,, 
\end{equation}
where $R$ is  the Fourier transform of $\sqrt{r(p,M)}$ given by
\begin{equation}
R(x-y,M)= \int \frac{d^4p}{(2\pi)^4}\,e^{ip\cdot(x-y)} \,\sqrt{r(p,M)} \,.
\end{equation}
Next, we promote $\phi(p,M)$ (and $a(p,M)$) as quantum operators, finally writing it in terms of the ladder operator we have,
\begin{equation}\label{decomposition}
\phi(x,M) = \frac{1}{(2\pi)^3}  \int \sqrt{ \frac{r(p,M)}{2 \omega_p}} \left[ a(\vec{p}) e^{-i p x} +   a^\dag (\vec{p}) e^{i p x}   \right] d^3 \vec{p} \,,
\end{equation}
where the standard dispersion relation is implemented into $r(p,M)$:
\begin{equation}\label{rr}
r(p,M) := r(p_0=\omega_p,\vec{p},M)   \,,
\end{equation}
with $\omega_p=\sqrt{\vec{p}^2+m^2}$. As it will be clear later, this is a necessary condition.

One implements the standard canonical commutation relations (CCR) but on the field $\phi(p)$ in \eqref{eq:phi_p_classic},
\begin{equation}\label{CCR}
[a(\vec{p}),a^\dag(\vec{p'})]= (2\pi)^3\,\delta(\vec{p}-\vec{p'})   \hspace{1em}   \Leftrightarrow   \hspace{1em}      [\phi(t,\vec{x}),\pi(t,\vec{y})]= i \delta(\vec{x}-\vec{y}) \,,
\end{equation}
where $\pi=\dot{\phi}$ is the conjugate variable of $\hat{\phi}$, and the square brackets denote the commutator. The  CCR can be rewritten as,
\begin{equation}\label{CCR2}
[a(\vec{p},M),a^\dag(\vec{p'},M)]= (2\pi)^3\,r (\vec{p},M)\delta(\vec{p}-\vec{p'}) =(2\pi)^3\, r (p,M)\delta(\vec{p}-\vec{p'})       \,, 
\end{equation}
where the last equality comes from \eqref{rr}. The action on the vacuum becomes 
\begin{equation}\label{on_zero}
a^\dag(\vec{p},M) |0\rangle = \sqrt{r(\vec{p},M)} |1_{\vec{p}} \rangle    \hspace{2em}   a(\vec{p},M) |0\rangle = 0 \,. 
\end{equation}
Therefore, the action of the field on the vacuum leads to a plane wave but with a scale-dependent prefactor:
\begin{equation}\label{on_zero2}
\phi(x,M) |0\rangle =\frac{1}{(2\pi)^3} \int \frac{d^3 p}{\sqrt{2 \omega_p}} \sqrt{r(\vec{p},M)} e^{i p x} |1_{\vec{p}} \rangle    \,. 
\end{equation}
The generalization of the quantization, in the presence of dimensional reduction, of vector and fermion fields is straightforward. It is
sufficient to decompose the fields as in \eqref{eq:phi_p_classic} and then write covariant CCR, for vector fields, and anti-commutation relations (ACR) for fermions. The Reader might worry about the quantization of vector fields, then of gauge bosons, in the presence of a dimensional scale ($M$) since this may resemble a cutoff, thus inconsistent with gauge invariance. However, this is not the case. A sharp cutoff rules out the high momentum modes breaking the gauge invariance; conversely, the Fourier transforms of the fields, e.g., \eqref{decomposition}, range up to infinite energy.

\subsection{Propagator}

Generalizing \eqref{standard_Fey}, the Feynman propagator is
\begin{equation}\label{nonstandard_Fey}
G_F(x-y,M) = \langle 0 | \mathcal{T} \phi(x,M) \phi(y,M)  | 0 \rangle  \,.
\end{equation}
Replacing the field with the representation in \eqref{decomposition} and using \eqref{on_zero2} gives,
\begin{equation}\label{nonstandard_Fey_2}
G_F(x-y,M) = \frac{1}{(2\pi)^3}\int d^3 \vec{p} \frac{r(\vec{p},M)}{2 \omega_p} e^{-i p(x-y)} \,,
\end{equation}
which has to be written in terms of 4-vectors. We achieve this in the usual way, but with some attention to the function $r$. We call for shortness $s:=x-y$ and consider the integral, 
\begin{equation}
\frac{1}{(2\pi)^4} \int d^3 \vec{p} e^{i \vec{p} \, \vec{s}}  \int d p_0 \frac{i r(p_0,\vec{p},M) e^{-i s_0 \, p_0}}{p^2-m^2} \,.
\end{equation}
Writing $p^2-m^2= (p-\omega_p)(p+\omega_p) $, integration on the known Feynman contour, and picking up the pole at $\omega_p -i \epsilon$ leads to,
\begin{equation}
\frac{1}{(2\pi)^3} \int d^3 \vec{p} \frac{r(p_0=\omega_p,\vec{p},M)}{2 \omega_p} e^{-i p(x-y)} \,. 
\end{equation}
Comparing this with \eqref{nonstandard_Fey_2} implies,
\begin{equation}\label{rrbis}
r(\vec{p},M) = r(p_0=\omega_p,\vec{p},M) \,,
\end{equation}
which is $r(p,M)$ due to \eqref{rr}. Therefore, the propagator in momentum space, expressed through 4-vectors, is
\begin{equation}\label{nonstandard_Fey_3}
G_F(p,M) = i \frac{r(p,M)}{p^2-m^2 + i \epsilon} \,. 
\end{equation}
It is important to elaborate on the implications of \eqref{rr} and \eqref{rrbis}.
By comparing \eqref{new_prop} and \eqref{nonstandard_Fey_3} (regardless that the latter is in Minkowski space) , one sees that $r$ is an even function in momentum. This will turn out to be problematic for the consistency of the theory. The function $r$ must be asymmetric for $p\to - p$ (we shall refer to it simply as odd), so we can modify it as
\begin{equation}\label{r_form}
r(p,M,v) = V\, e^{-(\frac{p}{M}-v)^2}
\end{equation}
where $v$ is a constant 4-vector necessary to build $r$ as a function not only of $p^2$, and $V=\exp[v^2]$ is a normalization constant.
This amounts to a redefinition of the form factor in momentum space in \eqref{newGp}, namely, shifting $p\to p-M\,v$.

Among others, the introduction of a special direction, $v$, will imply the breaking of the rotational invariance of the vacuum. Even more important for the rest of this work, the form of the function $r$ in \eqref{r_form}, odd in the variable $p$, will break the spatial translational invariance of the vacuum, with relevant impact for the consistency of the theory, discussed in section \ref{Consistency}.

However, the insight on dimensional reduction from  \eqref{new_prop} remains unchanged since it is only based on arguments of convergence or divergence of the propagator, dominated by the $p^2$ contribution, which reflects on the loop behavior or the couplings running.  
Notice also that $r$, in the form of \eqref{r_form}, does not trivialize on-shell, yielding effects even for tree-level processes.

\subsection{Translations, rotations, and the vacuum}

Let us first focus on the spatial translations, which, as anticipated, play a central role in the proposed framework.

Similarly to the standard case, the 3-momentum operator is given by \footnote{The vacuum-to-vacuum expectation value of $P_i$ is zero in the standard QFT, but \textit{a priori} not the one of $P_0$. Often, this is artificially set to zero by Normal Product. One reason is that the corresponding integral rapidly diverges. This is in contrast with the QFT in the multifractal spacetime, where the finiteness does not require artifacts in the regularization.}
\begin{equation}\label{intP}
P_i = \int d^3 \vec{x} \, \dot{\phi}(x,M) \partial_i \phi(x,M) = \frac{1}{(2\pi)^3} \int d^3 \vec{p} \, \frac{p_i}{2} \left( a^\dagger(\bar{p}, M) a(\bar{p}, M) + a(\bar{p}, M) a^\dagger(\bar{p}, M) \right)\,,
\end{equation}
being the dot the temporal derivative. 

By evaluating the vacuum expectation value with the help of equation \eqref{on_zero}, one finds (for shortness, we indicate $r(p,M,v)$  just as $r(p)$)
\begin{equation}\label{no_trans_zero}
\langle 0 | P_i | 0 \rangle = \frac{1}{(2\pi)^3} \int d^3 \vec{p} \, \frac{p_i}{2} r(\vec{p}) \,,
\end{equation}
which generalizes the standard expression, here modified by $r$.

Due to the form of the function $r$ in equation \eqref{r_form}, and in contrast with the standard QFT, this integral is non-zero but finite. In other words,
\begin{equation}\label{no_trans}
P_i | 0 \rangle \neq  0 \,,
\end{equation}
namely, the vacuum possesses momentum. The $P_i$ are the generators of the spatial translation, $T=\exp[i\, b_j P_j]$, being $b_j$ the translational parameters, so the vacuum is not translational invariant:
\begin{equation}\label{no_trans2}
T | 0 \rangle \neq | 0 \rangle   \,. 
\end{equation}
Since rotations are written as volume integrals of the type in equation \eqref{intP}, where one takes the textbook expressions and replaces $\phi(x) \to \phi(x,M)$ and then $a(\vec{p}) \to a(\vec{p},M)$, the analog conclusion of equation \eqref{no_trans} also holds for rotations. Thus, the theory, due to the deformation of the generators of translation or rotations caused by $r$, shows a breaking at the quantum level of the Poincar\'e invariance.

The linear term $p \cdot v$ in the exponent of \eqref{r_form} represents a crucial departure from the form factors typically employed in nonlocal quantum gravity \cite{Modesto:2012prd,Tomboulis:1997,Biswas:2012prl}. Standard nonlocal theories utilize form factors of the type $\exp(-p^2/M^2)$, which depend only on $p^2$ and therefore preserve the Poincaré invariance of the vacuum. In contrast, our form factor explicitly breaks translational invariance through the $(p/M - v)$ structure, where $v$ is a constant four-vector that selects a preferred frame. This distinction is not merely technical but conceptually fundamental: while both approaches introduce exponential UV suppression, only the breaking of vacuum translational invariance—encoded in the $v$-dependence—circumvents Haag's theorem \eqref{no_trans}, \eqref{no_trans2} and enables consistent construction of the interaction picture, as will be discussed in Sec.\ref{Consistency}. Nonetheless, standard nonlocal theories achieve UV finiteness and improved renormalization properties but remain subject to Haag's constraints; our framework sacrifices exact Poincaré invariance of the vacuum to resolve the more fundamental obstruction to perturbative QFT posed by Haag's theorem. The physical interpretation is that the multifractal structure of spacetime itself singles out a preferred direction encoded in $v$, with observable consequences in high-energy scattering cross-sections that vanish in the low-energy limit $p \ll M$.

\section{Consistency of the theory}\label{Consistency}

In this section, our proposal confronts two milestones of QFT: the Källén-Lehmann representation of the propagator; the Gell-Mann and Low theorem for the propagator -- once an interaction, say $\frac{\lambda}{4!}\, \phi^4$, is turned on. 

While the nonperturbative Källén-Lehmann representation bounds the behavior of the propagator, thus providing a consistency check for the theory, the  Gell-Mann and Low formula, base for perturbation theory, is not a test for the theory but is an improved result within nonstandard QFT, including the scale $M$.

Notice that both these well-known results are related to the Poincar\'e (non)invariance of the vacuum. We shall now go through the issue in detail.

\paragraph{Källén-Lehmann spectral representation.}

Recent work has shown that nonlocal theories with entire form factors and a Poincaré-invariant vacuum admit a generalized Källén–Lehmann representation with a positive spectral density and the correct local limit \cite{Briscese:2024jhep}. Our scenario is complementary: because the vacuum breaks translational invariance, a key step in the standard derivation fails. The usual KL bound does not apply to the discussed scenario.

In standard QFT, the Källén-Lehmann representation of the propagator reads,
\begin{equation}\label{KL}
G(p)= \int_0^\infty d(\mu^2) \rho(\mu^2) \frac{1}{p^2-\mu^2}\,,
\end{equation}
where $\rho$ is a positive definite function, known as the spectral function.

This equation implies that the propagator cannot decay faster than $p^{-2}$ for large momentum, in apparent contrast with \eqref{nonstandard_Fey_3}. However, we argue that the Källén-Lehmann representation does not hold in the multifractal QFT, thereby invalidating the constraint. The key point is that deriving the Källén-Lehmann representation in \eqref{KL} requires the Poincar\'e invariance of the vacuum to manipulate the expression $\langle\Omega| \phi(x) \phi(y)| \Omega\rangle $.

In particular, one has in the fractal  QFT,
\begin{equation} \label{KLvacuum}
\langle \Omega | \phi(x) \phi(y) | \Omega \rangle  = \sum_n \langle \Omega | \phi(x) | n \rangle \langle n |\phi(y) | \Omega \rangle = e^{i p_{\Omega} (x-y)} \sum_n  e^{-i p_n (x-y)}
|\langle \Omega | \phi(0) | n \rangle|^2 \,, 
\end{equation}
In the first step, we used the completeness relation over a basis of four-momentum eigenstates $|n\rangle$;
in the second, we incorporated the non-translational invariance of the vacuum via the vacuum momentum $p_\Omega$. The crucial point is that $p_\Omega$ is infinite since the vacuum carries a finite momentum density (see \eqref{no_trans_zero} and \eqref{no_trans}).
Therefore, the right-hand side of the expression in \eqref{KLvacuum} becomes ill-defined, implying that the Källén-Lehmann representation of the two-point function breaks down in multifractal QFT. Consequently, the standard constraint that the propagator cannot decay faster than $1/p^2$ at large $p$ no longer applies.

\paragraph{Gell-Mann and Low theorem and S-matrix construction.}

While the previous paragraph is a consistency check, this section demonstrates how fractal QFT solves some fundamental issues in standard QFT through Poincar\'e non-invariance of the vacuum. This property, shown in equation \eqref{no_trans}, enables the construction of a consistent $S$-matrix and proper application of the Gell-Mann and Low formula since broken translational invariance invalidates Haag's theorem \cite{Haag:1955ev}. Standard QFT suffers from Haag's theorem \cite{Haag:1955ev}, which prohibits a non-trivial interaction picture or perturbation theory -- a no-go result often overlooked in contemporary literature.
 
Let us formulate this precisely. We denote the vacuum states in interactive and non-interactive Fock spaces as $|\Omega \rangle$ and $|0 \rangle$, respectively, and represent interactive and non-interactive fields as $\phi$ and $\phi_0$. The evolution operator relating free and interacting theory for $t > t'$ is:
\begin{equation}
U(t,t')= \exp\left[-i \int_{t'}^t H_I(\tau) d\tau \right]
\end{equation}
Standard QFT assumes the existence of the unitary operator connecting the free and interacting fields through:
\begin{equation}\label{noU}
\phi(t,\vec{x}) = U^\dagger(t,t_0) \phi_0(t,\vec{x}) U(t,t_0) \,.
\end{equation}
Moreover, the free and interactive vacua are related as (calling $t'=0$ and $t=T$, as $T \rightarrow \infty$) \cite{Gell-Mann:1951ooy},
\begin{equation}\label{noU2}
|\Omega\rangle = \frac{U(0,\pm T)|0\rangle}{\langle 0| U(0,\pm T) |0\rangle} \,.
\end{equation}
Plugging \eqref{noU2} into $\langle \Omega | \mathcal{T} \phi(x) \phi(y) |\Omega \rangle$ and performing simple algebra yields the Gell-Mann and Low relation:
\begin{equation}\label{GL}
\langle \Omega | \mathcal{T} \phi(x) \phi(y) |\Omega \rangle = \frac{\langle 0 | \mathcal{T} \phi_0(x) \phi_0(y) \exp[-i \int_{-\infty}^\infty H_I(t) dt] | 0 \rangle}{\langle 0 | \mathcal{T} \exp[-i \int_{-\infty}^\infty H_I(t) dt] | 0 \rangle}
\end{equation}
Unfortunately, Haag's theorem establishes that equations \eqref{noU} and \eqref{noU2} are valid only in the free-field case, namely, when $U$ is the identity -- meaning that the interaction picture exists only when no interaction exists \footnote{For an overview of the interpretations of Haag's theorem, see \cite{Mitsch:2022oyw}.}.

For the present discussion, the crucial point is that fractal QFT circumvents this issue through its non-translational invariant vacuum. The proof of Haag's theorem depends on vacuum translational invariance (see appendix \ref{HT}), a condition explicitly violated in fractal QFT as shown in equation \eqref{no_trans}. This violation enables fractal QFT to construct a consistent operator $U$ in \eqref{noU} and \eqref{noU2}, while the logical steps to arrive at \eqref{GL} remain unchanged. As a result, fractal QFT guarantees a robust perturbative framework, remarkably improving the standard theory -- see also section \ref{Discussion}.

\section{Possible Implications and Discussion}\label{Discussion}

In this section, we explore some implications of fractal QFT. The most striking consequence of the adopted form factor is perturbative finiteness. All loops become convergent, and the running coupling ceases to grow logarithmically at high energies, eliminating UV renormalons. Recall that renormalons lack any semi-classical interpretation that drives the factorial ($n!$) divergence in renormalized perturbation theory \cite{tHooft:1977xjm}.

\paragraph{Loop finiteness.}

The appearance of the function $r$ in \eqref{nonstandard_Fey_3} renders loop integrals finite, creating a perturbatively finite theory. Consider the 4-point correlator in the $\phi^4$ model with interaction Lagrangian $\frac{\lambda}{4!} \phi^4$—the so-called "fish-diagram." This becomes finite due to the suppression from $r$ at high momentum. Since $r(p)\approx 1$ for $p\ll M$ while providing exponential suppression for $p> M$, the integral behaves as $\log(p^2/\mu_0^2)$ for $p\ll M$, and as $M^2/p^2 \exp (-p^2/M^2)$. Consequently, the one-loop running becomes:
\begin{equation}\label{phi4_run}
\left\{
\begin{array}{ll}
\lambda(\mu) \simeq \lambda(\mu_0)+ \left(  1+  \lambda(\mu_0) \beta_1 \log(\mu^2/\mu_0^2) \right) &  \mu < M     \\[10pt]
\lambda(\mu) \simeq \lambda(M) \left(1 - \lambda(M) \beta_1 \frac{M^4}{4\mu^4} e^{-\frac{\mu^4}{M^4}} \right) &  \mu \geq M    \,, \\
\end{array}
\right.
\end{equation}
where the usual one-loop factor is $\beta_1=3/(16\pi^2)$. The first line of \eqref{phi4_run} shows standard running; the second reveals that the coupling quickly reaches an asymptotic constant value for $\mu \geq M$.

In summary, any loop constructed in the model becomes finite, and no Landau pole exists due to \eqref{phi4_run}. This points to super-renormalizability or even finiteness, properties extensively studied in nonlocal quantum gravity \cite{Modesto:2012prd,Tomboulis:2015prd,Modesto:2017review,Calcagni:2022shb,Calcagni:2023spectral}. The exponential form factor suppresses UV divergences at all loop orders.

\paragraph{Anisotropies in high-energy scatterings.} We should also examine modifications to on-shell processes, such as $2\rightarrow 2$ scattering. At leading order in $\lambda$ expansion, we have:
\begin{align}
& \langle f | S^{(1)} | i \rangle = \langle 1_p, 1_{p'} | (-i\, \lambda) \, \phi^-(x) \phi^-(x) \phi^+(x) \phi^+(x) | 1_k, 1_{k'} \rangle = \nonumber \\
& \frac{1}{(2\pi)^8} (-i\, \lambda) \, \delta^{(4)}(p+p'-k-k')\, \frac{\sqrt{r(p)\, r(p')\, r(k)\, r(k')}}{\sqrt{(2\omega_{\vec{p}})\,  (2\omega_{\vec{p}'})\, (2\omega_{\vec{k}})\, (2\omega_{\vec{k}'})}} \label{amplitude}   \,.
\end{align}
The superscripts $-$ and $+$ denote the creation and destruction operator parts of the field.
The standard result is modified only by the factors $r$ in \eqref{amplitude}.
Given the form of the function $r$ in \eqref{r_form}, the part in $p^2$ becomes trivial on-shell, $p^2=m^2$, while the part in $p_\mu v^\mu$ produces an anisotropy. Some momentum directions may be enhanced, depending on the constant four-vector $v$. Of course, all these modifications are suppressed at energies much lower than $M$.

\paragraph{Absence of the UV renormalons.}

We have discussed the perturbative finiteness of the theory above; now, let us consider what to expect at the non-perturbative level.  Renormalons \cite{tHooft:1977xjm,Parisi:1978iq} can be interpreted as a bridge between perturbative and non-perturbative physics. The singularities in the Borel transform due to renormalons lie on the semi-positive axis and hamper any Borel-Laplace resummation. They are not related to any semi-classical expansion and are considered genuine non-perturbative objects related to renormalization \cite{Parisi:1978iq}. In the literature, their presence represents a failure of the perturbative renormalization procedure since the resulting series cannot be resummed without ambiguities \cite{tHooft:1977xjm}.

Returning to the $\phi(x)^4$ model, renormalons relate to factorially divergent series\footnote{Independently of the instantons that, notwithstanding leading to a $n!$ large order behavior of the perturbation theory series, their existence does not imply any inconsistency in the theory, as explained in \cite{tHooft:1977xjm}. } obtained by evaluating Feynman diagrams with many insertions of the same sub-diagram (the fish-diagram). The origin of the factorial $n!$ (where $n$ is the perturbation theory order) stems from the logarithmic high-energy behavior of the fish-diagram (sub-diagram) inserted $n$ times. While we refer to \cite{Beneke_1999} for a complete derivation of the $n!$ renormalon behavior (see \cite{Bersini_2020} for an alternative approach), here we illustrate its origin with a simple example. Consider the integral:
\begin{equation}\label{sketch}
\int_{\mu_0}^\mu \log(\frac{p}{\mu_0})^n d^4p \,.
\end{equation}
This expression mimics a so-called bubble diagram, i.e., a loop integral with $n$ nested occurrences of the
4-point correlator (fish diagram), which has logarithmic UV behavior in standard QFT.

Renormalons are calculated from the finite part of the loop, so focusing on the finite part of \eqref{sketch} as $\mu\rightarrow \infty$, we obtain:
\begin{equation}
\left(- \frac{1}{4} \right)^{1+n} \mu_0^4 \, n!  \,.
\end{equation}
This suffices to trace the factorial divergence of the series due to renormalons back to the logarithmic UV behavior of the running coupling.

Here lies the improvement provided by dimensional reduction theory. Due to \eqref{phi4_run}, the behavior for $\mu\rightarrow \infty$ is not logarithmic, unlike standard QFT. This suffices to eliminate the renormalon ambiguities.

It is important to emphasize that the absence of UV renormalons and the absence of the UV Landau pole are distinct phenomena that need not occur together. Typically, one might encounter situations where renormalons emerge from the perturbative evaluation of bubble diagrams even when the Landau pole is absent due to nonperturbative dynamics, as discussed in \cite{Maiezza_2021,Maiezza_2024}. However, our current framework provides a more comprehensive solution: within the perturbative approach itself, both the UV Landau pole and renormalons simultaneously disappear, highlighting the theoretical consistency of our model. 

A comment is finally in order. In \cite{Maiezza:2020qib}, we conjecture that since UV renormalons indicate a failure of self-consistent perturbative renormalization, they might be reinterpreted as ignorance of the no-go provided by the Haag theorem. Standard perturbative divergences might also signal a warning about this no-go, though perturbative renormalization addresses this problem. However, renormalized perturbation theory needs resummation for self-consistency, which renormalon ambiguities prevent. The present work supports this interpretation: we show that when the Haag theorem's no-go disappears, so do renormalons (and perturbative divergences).

\paragraph{Implications.}

Within the multifractal spacetime, the standard model becomes automatically  finite at any order in perturbation theory and avoids UV Landau poles. Consequently, it achieves asymptotic safety and remains valid at all scales—see \cite{Litim_2014} for an in-depth discussion on the asymptotic safety paradigm in QFT. For example, figure \ref{plotSM} illustrates the one-loop running of the standard model's gauge couplings.

It's worth commenting on the nature of safety shown in Fig. \ref{plotSM}. Wilson first argued for the necessity of a non-trivial UV fixed point for theoretical consistency \cite{Wilson:1971bg}. In our case -- QFT with diminishing dimension -- the UV fixed point is reached only at infinity. Nevertheless, the couplings remain finite and possibly small. While the fixed point is attained at infinite energy, the running exponentially converges to a constant value. 

The realization of asymptotic safety discussed in this work differs fundamentally from Weinberg's asymptotic safety program \cite{Weinberg:1979,Reuter:1996,Lauscher:2001,Percacci:2009,Reuter:2012book} and its implementations in functional renormalization group (FRG) approaches to quantum gravity \cite{Niedermaier:2006,Codello:2008,Eichhorn:2018review}. In standard asymptotic safety, UV finiteness arises from a non-Gaussian fixed point of the gravitational RG flow where dimensionless couplings become scale-independent, with dimensional reduction emerging as a consequence of the fixed-point structure \cite{Lauscher:2005,Reuter:2012spectral}. The theory remains perturbatively non-renormalizable but is well-defined non-perturbatively through the fixed point, preserving Poincaré invariance throughout. In contrast, our framework achieves asymptotic safety through the exponential form factor suppression $e^{-p^2/M^2}$ in \eqref{r_form}, which causes the running coupling to asymptotically approach a constant \eqref{phi4_run} without reaching a genuine RG fixed point—the coupling continues to vary, albeit exponentially slowly, rather than becoming exactly scale-independent. Moreover, our mechanism operates entirely within an improved perturbation theory (due to Haag's no-go evasion and a coupling that can remain small during the running). While both approaches point to dimensional reduction at high energies, the physical mechanisms differ. The standard asymptotic safety relies on strong quantum gravity fluctuations driving the theory to a fixed point, whereas the approach presented here accomplishes the UV completion through the form factor, effectively parameterizing the flow of the spacetime dimension.

As illustrated in \eqref{amplitude}, the multifractal QFT predicts direction-dependent cross-sections. Phenomenologically, within the standard model, this would probably be the most striking ``smoking gun" signal of the underlying space-time fractality. This breaking of the vacuum's translational and rotational invariance, encoded in the vector $v$, implies that Lorentz invariance is also broken at a fundamental level. This is a crucial feature of our model with significant phenomenological implications. While a detailed analysis of experimental bounds is beyond the scope of this work \cite{Dreissen_2022}, it is essential to address a key theoretical consistency check for such theories.

A critical challenge for Lorentz-violating theories is the potential for radiative corrections to amplify symmetry-breaking effects to phenomenologically unacceptable levels \cite{Collins_2004,collins2006lorentzinvarianceviolationrole}. Crucially, our framework is protected from this instability due to the term $e^{-p^2/M^2}$ in the form factor of \eqref{r_form}. The amplification mechanism is a direct consequence of having a form factor of the form $F(p\cdot v)$, with $F(p\cdot v)=1$ for $p\rightarrow 0$ and $F(p\cdot v)=0$ for $p\rightarrow \infty$. In our model, the form factor in \eqref{r_form} ensures renormabizability is preserved, namely unsupressed contributions proportional to positive powers of the scale $M$ can be reabsorbed into a redefinition of the couplings present in the tree level Lagrangian. Thus  any Lorentz-violating  contribution must be suppressed by negative powers of the large scale $M$. This eliminates the source of any pathological enhancement and ensures decoupling. See Appendix \ref{sec:quantum_stability} for a detailed and explicit discussion.

 \begin{figure}[t]
 \centerline{
\includegraphics[width=.65\columnwidth]{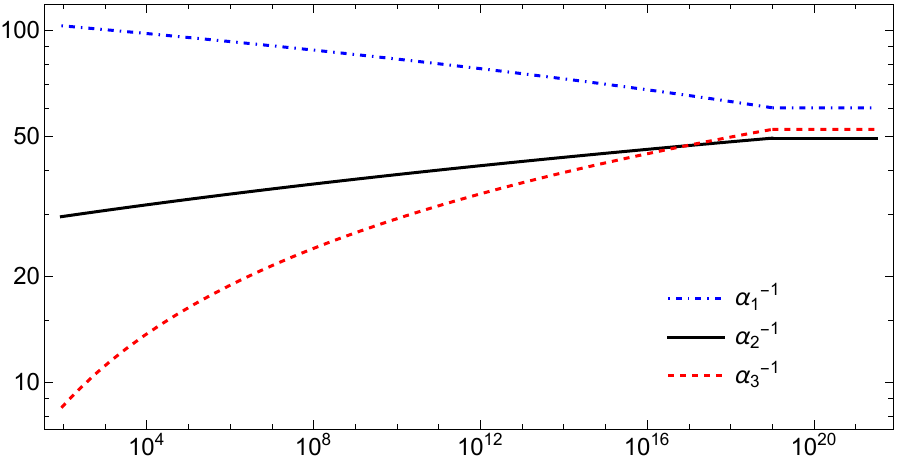}
}
\caption{Running of the gauge couplings of the standard model, which asymptotically reach constant values.}
\label{plotSM}
\end{figure}

\section{Conclusions and outlook}\label{Summary}


Motivated by the possibility that spacetime's effective dimension may evolve with energy scale, we have explored this concept from a quantum field theory perspective, particularly within canonical quantization. Crucially, consistency of the quantum field theory points to a quantum nature of spacetime, with classical smoothness emerging only as a low-energy approximation.

We have shown how to quantize fields in a spacetime with multifractal properties and how this automatically implies dimensional reduction, ensuring consistency with established field theory constraints. Notably, not only is quantization compatible with a varying dimension, but such variation significantly enhances QFT's robustness, enabling rigorous $S$-matrix construction in the interaction picture (perturbation theory). This results in a finite theory without loop divergences and improves the perturbative series' behavior. The significance of our contribution lies in demonstrating that non-differentiable spacetime structures are not only compatible with quantum field theory but also actively improve its mathematical foundation. All this is achieved while retaining all known behaviors and predictions of standard QFT at low energies, yet predicting new behaviors at large energies, mainly through ``asymptotic safety" and small anisotropies at high-energy scatterings. 

When comparing our work with existing literature, some apparent similarities arise between our quantization in section \ref{sec:Quantization} and the works in \cite{Altaisky:2006dj,Altaisky:2010wv}, where the author proposes scale-dependent QFT. However, the concept of scale dependence in those works has no connection to our fractality and dimensional reduction. The conceptual framework, technical details, and outcomes are fundamentally different. Some analogy in intent exists with our earlier proposal \cite{Maiezza:2022xqv}. However, important differences exist concerning what we proposed in this work. The approach in \cite{Maiezza:2022xqv} requires modifications due to defining the action via Stieltjes integrals, entailing modifications to the equations of motion that must be addressed with certain constraints. The Stieltjes-integral action approach in \cite{Maiezza:2022xqv} is  more technically involved and becomes inconsistent when introducing Lagrangians with additional fields, particularly fermions. Conversely, our present approach requires only a specific quantization of the standard action, incorporating a dimensional reduction scale, and it remains unaltered with the addition of new fields, fermions included. A radically different route to model fractal spaces is the one in \cite{calcagni2013geometryfractionalspaces}, based on fractional calculus. A final parallel can be drawn with \cite{Shirkov_2010,Fiziev:2010je}; however, these references attempt to address varying topological dimensions, rendering the framework genuinely distinct.

Looking beyond particle physics, we should also comment on the possible impact of \eqref{r_form} on the standard cosmological model. First, the vacuum energy density becomes finite, like any integral in the theory, but remains remarkably large -- approximately $O(M^4)$. Standard QFT requires regularization and renormalization to control divergences; conversely, in fractal space-time QFT, integrals are inherently finite, though the renormalization group equation remains applicable. This property is generally independent of divergences. Thus, even in a universe with running dimensions, QFT does not produce a small cosmological constant. The smallness must instead be understood within the renormalization group framework, where parameter values are in principle arbitrary and cannot be predicted.

Second, and more intriguingly, the introduction of a specific direction -- represented by vector $v$ in \eqref{r_form} -- indicates a fundamental anisotropy of space-time. As highlighted in \eqref{no_trans}, in QFT with dimensional reduction, the vacuum possesses not only energy but also momentum along a particular direction. Consequently, even though all the possible new effects are suppressed by the scale $M$, the stress-energy tensor is expected to develop non-diagonal components. This might impact the Friedmann equation, potentially suggesting non-standard cosmological effects, namely, departing from the cosmological principle \cite{Adam_2025}. Further exploration of this direction is compelling but beyond our current scope.

In summary, while phenomenological studies in the literature investigate potential effects of varying dimensions \cite{Anchordoqui:2010hi,Mureika:2011bv,ANCHORDOQUI_2012,Stojkovic:2013xcj}, we believe our work provides the theoretical foundation for such scenarios. A dedicated study will be needed to analyze further consequences for particle physics and cosmology.


\section*{Acknowledgement}

AM thanks Emanuela Pichelli for discussions during this work on the appearance of fractals in the physical world. JCV give special thanks to Paolo Creminelli , Marco Serone, and Andrea Romanino for support. JCV also gives special thanks to SISSA International Relations Office, in particular to Francesca Gandolfo and Gabriella Perissini for  help and guidance. 


\appendix

\section{S-matrix and Poincar\'e invariance of the vacuum}\label{HT}

This appendix highlights some known problems in building a consistent $S-$matrix in the interaction picture in QFT.
 
Equation \eqref{GL} relies on the interaction picture where the free field, $\phi_0$, and the interactive field, $\phi$, are unitarly related:
\begin{equation}\label{unitary}
\phi(t,\vec{x})= U(t_0,t)^\dag \phi_0 (t,\vec{x}) U(t_0,t) \,,
\end{equation}
and the vacua $|0\rangle$ and  $|\Omega\rangle$ of the free and interactive Fock spaces, respectively, are two distinct vectors, e.g., as \eqref{noU2},
%
\begin{equation}\label{noeq}
| \Omega \rangle  \neq | 0 \rangle    \,.
\end{equation}
In standard QFT, namely, in the absence of fractality and reduction scale, the unitary operator $U(t_0,t)$ exists only in the trivial case, with no interaction, for which such an operator is just the identity. This result is known as the Haag theorem, heavily relying on the translational invariance of the vacuum. It may be helpful to recall how this invariance leads to the no-go of the theorem, following \cite{Maiezza:2020qib} -- see \cite{hall1957theorem} for a rigorous proof within axiomatic QFT. 

For shortness, we denote $U(t_0,t):= U$ and call $T_0$ and $T$ the translational operators in the free and interactive Fock spaces. The translational invariance of the vacua reads,
\begin{equation}\label{tr_inv}
T_0 |0\rangle = |0\rangle \hspace{2em}       T |\Omega\rangle = |\Omega\rangle \,.
\end{equation}
Calling $\vec{b}$ the translational parameter, the operators $T_0$ and $T$ act on the fields as,
\begin{equation}\label{Tonphi}
T_0^\dag \phi_0(t,\vec{x}) T_0 = \phi_0(t,\vec{x}-\vec{b})   \hspace{2em}   T^\dag \phi(t,\vec{x}) T = \phi(t,\vec{x}-\vec{b}) 
\end{equation}
From the latter and \eqref{unitary}, we have,
\begin{equation}\label{side1}
T^\dag \phi(t,\vec{x}) T = \phi(t,\vec{x}-\vec{b}) = U^\dag \phi_0(t,\vec{x}-\vec{b}) U =  U^\dag T_0^\dag \phi_0(t,\vec{x}) T_0 U \,.
\end{equation}
On the other hand, we have,
\begin{equation}\label{side2}
T^\dag \phi(t,\vec{x}) T = T^\dag U^\dag \phi_0(t,\vec{x}) U T \,.
\end{equation}
Comparing \eqref{side1} and \eqref{side2} yields,
\begin{equation}\label{UT}
U T = T_0 U    \,.
\end{equation}
Next, multiply from the right \eqref{UT} for $|\Omega\rangle$,
\begin{equation}
U T |\Omega\rangle = U |\Omega\rangle= T_0 U |\Omega\rangle   \,.
\end{equation}
and comparing the last equality with the first of \eqref{tr_inv} (i.e. $T_0 |0 \rangle = |0 \rangle$) gives
\begin{equation}\label{UtoV1}
U |\Omega\rangle = |0 \rangle \,,
\end{equation}
or equivalently
\begin{equation}\label{UtoV2}
|\Omega\rangle = U^\dag |0 \rangle \,.
\end{equation}
These equations conflict with \eqref{noU2}.

Moreover, from \eqref{UtoV1} and \eqref{UtoV2}, we can write,
\begin{equation}
\langle \Omega | U | \Omega \rangle = \langle\Omega|0\rangle   \hspace{2em}  \langle 0 | U^\dag | 0 \rangle = \langle 0|\Omega \rangle\,.
\end{equation}
Finally, these two equations lead to the following chain of equalities,
\begin{equation}
\langle \Omega | U | \Omega \rangle = \langle\Omega|0\rangle= \left( \langle 0|\Omega \rangle \right)^\dag = \langle 0 | U | 0 \rangle\,,
\end{equation}
implying (modulo an irrelevant overall phase)
\begin{equation}
| \Omega \rangle = | 0 \rangle    \,,
\end{equation}
in contradiction with \eqref{noeq}, unless the unitary operator $U$ coincides with the identity. This case, however, is the one with no interactions, hence the Haag theorem. Since the $S-$matrix is $U(-\infty,\infty)$, the Haag theorem is a no-go for perturbative QFT. The multifractality and dimensional reduction dramatically change \eqref{tr_inv}, enabling the construction of a consistent $S-$matrix.

\section{Quantum corrections do not enhance Lorentz violation}\label{sec:quantum_stability}
In this appendix, we argue that the framework presented in this paper avoids the pathologies pointed out in \cite{Collins_2004,collins2006lorentzinvarianceviolationrole} regarding loop-enhanced Lorentz asymmetries that rule out the phenomenological viability of some models. 

Consider the one-loop correction to the self-energy $\Pi$ in the $\phi^4$ model, embedded in the setting of this work, namely with
the propagator modified by the form factor in \eqref{r_form}. The one-loop contribution reads (in Euclidean momentum),
\begin{equation}\label{example_expl}
\Pi= \frac{\lambda}{2} \int \frac{d^4 l_E}{(2\pi)^4} \frac{e^{-l_E^2/M^2-l\cdot v/M}}{l_E^2+m^2} \,.
\end{equation}
Since we are interested in the UV behavior, we evaluate the integral in the limit $m=0$. We write $l\cdot v= |l|\times |v| \cos(\theta)$ and rewrite the 4D Euclidean integration as
\begin{equation}
\int_0^\infty  \int_0^\pi \int_0^\pi \int_0^{2\pi} \frac{e^{-l_E^2/M^2-l\cdot v/M}}{l_E^2} l_E^3  \sin(\theta)^2 \sin(\varphi) dl_Ed\theta d\varphi d\omega \,.
\end{equation}
The momentum integration leads to:
\begin{equation}
\frac{1}{4} M^2 \left(2-\sqrt{\pi } v \cos (\theta ) e^{\frac{1}{4} v^2 \cos ^2(\theta )} \text{erfc}\left(\frac{1}{2}
   v \cos (\theta )\right)\right) \,.
\end{equation}
where erfc is the complementary error function.

Next, we proceed to the $\theta-$integration. Here comes the crucial point: upon integration over $\theta$, the term linear in $v\cdot \cos(\theta)$ vanishes due to odd parity. Therefore, only even power of $v$ can appear in the result. 

More explicitly, after the trivial integration in $\varphi$ and $\omega$, the final result can be approximated as 
\begin{equation}
\Pi \simeq \frac{\lambda  M^2 \left(v^4+12 v^2+96\right)}{3072 \pi ^2} + O\left(v^5\right)\,.
\end{equation}
Thus, we have $\Pi$, resulting in a large finite term (due to the presence of $M$) that can be reabsorbed into the standard counterterms, specifically the mass counterterm. Note that the form factor provides regularization, but renormalization is still at work, and then these terms do not lead to unsuppressed Lorentz violating terms.

\medskip

To make contact with \cite{Collins_2004,collins2006lorentzinvarianceviolationrole}, suppose we had a form factor as
$$
e^{-l \cdot v/M}\,,
$$
now without the $l^2/M^2$ suppression. In this case, the integral corresponding to the one in \eqref{example_expl} would not automatically be finite. To see this, rewrite $l_E=l_E^p+l_E^t$, with $l_E^p, l_E^t$ being the parallel and transverse components with respect to the vector $v$. As a result, only the integral over $l_E^p$ would be finite, while those over $l_E^t$ (3 dimensions) would require a cutoff since $v \cdot l_E^t=0$. This would then be a direction-dependent cutoff in 3 dimensions, similar to the one discussed in Sect. 1.3 of \cite{collins2006lorentzinvarianceviolationrole}. In this example, one would then have enhanced Lorentz violation via the Lorentz-asymmetric cutoff. Thus, the crucial role in our framework—unlike in \cite{Collins_2004,collins2006lorentzinvarianceviolationrole}—is played by the term $\exp(-p^2/M^2)$, which makes the loop finite in any direction in momentum space.

The arguments above apply in general. Consider a generic expression for a loop calculation of the two-point Green function ($p^2\ll M^2$):
\begin{align}
 \Delta\Pi (p,v,m=0,M)= & \int \prod_{\ell=1}^m d^4\,\ell_l \, \mathcal{I}(\ell_l,p,v,m,M) =   A\, M^2   + \nonumber \\
 & B\,  p^2 + C\,\left(\frac{p\cdot v}{M}\right)\,p^2 + \mathcal{O}\left(\frac{p^{n+2}}{M^n}\right) \, .
\end{align}
where $A$, $B$, $C$ are functions of  $\frac{p^2}{M^2}$ and $ \frac{v\cdot p}{M}$. Since the renormalization properties of the theory are preserved, the Lorentz-violating finite contributions due to $v$ are automatically suppressed by the large scale $M$, unlike the potentially large Lorentz-violating contributions pointed out in \cite{Collins_2004,collins2006lorentzinvarianceviolationrole}.

\bibliographystyle{jhep}
\bibliography{biblio}

\end{document}